**The eDiscovery Medicine Show**

Maura R. Grossman, J.D., Ph.D.* and Gordon V. Cormack, Ph.D.*

As recently as 100 years ago, harmful practices such as bloodletting were still advanced as sound medical practice by expert practitioners.[1] Bloodletting gradually fell into disfavor as a growing body of scientific evidence showed its ineffectiveness and demonstrated the effectiveness of various pharmaceuticals for the prevention and treatment of certain diseases. Basking in the reflected glory of such scientifically proven medicines, unscrupulous purveyors of magical elixirs promoted their wares using pseudo-scientific evidence and testimonials from quacks and charlatans, presented along with free entertainment. These medicine shows persisted until, among other things, the Food and Drug Administration was given the authority to prosecute unsubstantiated therapeutic claims in 1938.[2]

eDiscovery methods, like therapeutics, are amenable to scientific evaluation. But practitioners and their "experts," vendors, and clients often ignore empirical evidence, citing instead existing or past practice to justify, for example, culling electronically stored information ("ESI") using untested search terms, establishing neither their necessity nor their efficacy. Or, they use pseudo-science to promote various potions marketed as "Artificial Intelligence," "AI," "technology-assisted review," or "TAR." Or, they employ pseudo-science and various logical fallacies to impugn scientific studies that contradict their claims. Or they point to the oft-cited Sedona Principle 6[3] as justification to do whatever they please. Or, sometimes, even all of the above. Trade shows and other "educational" activities sponsored by vendors promote their wares, complete with pseudo-scientific results, testimonials, sponsored receptions, prizes, and hospitality suites. The Continuing Legal Education ("CLE") industry and the trade press often echo these testimonials, failing to discriminate between practice and sound practice—let alone best practice—or between science and pseudo-science. So far, neither the courts nor any other authority has taken up the mantle, leaving parties to fend for themselves in the eDiscovery Wild West.

---

* Maura R. Grossman, J.D., Ph.D., is a Research Professor, and Gordon V. Cormack, Ph.D., is a Professor, in the David R. Cheriton School of Computer Science at the University of Waterloo, in Ontario, Canada. Professor Grossman is also Principal of Maura Grossman Law, an eDiscovery law and consulting firm in Buffalo, New York, U.S.A. Professor Grossman's work is supported, in part, by the National Sciences and Engineering Council of Canada ("NSERC"). The opinions expressed in this piece are the authors' own and do not necessarily reflect the views of the institutions, organizations, or clients with which they are affiliated. The authors wish to thank Jason R. Baron for his thoughtful comments on earlier drafts of this article.

[1] *See* William Osler, *The principles and practice of medicine: designed for the use of practitioners and students of medicine* (D. Appleton, 1892). *But see* Charles S. Bryan, *New observations support William Osler's rationale for systemic bloodletting*, in 32 Baylor U. Med. Ctr. Proc. 372-76 (Taylor & Francis, 2019), https://www.tandfonline.com/doi/full/10.1080/08998280.2019.1615331 (arguing that Osler's prescribed indications might be considered rational within the context of 19th century medicine).

[2] *See* Federal Food, Drug, and Cosmetic Act (June 24, 1938).

[3] Sedona Principle 6 states that "[r]esponding parties are best situated to evaluate the procedures, methodologies, and technologies appropriate for preserving and producing their own electronically stored information." The Sedona Conference, *The Sedona Conference Principles, Third Edition: Best Practices, Recommendations & Principles for Addressing Electronic Document Production*, 19 Sedona Conf. J. 1, 118-30 (2018), https://thesedonaconference.org/sites/default/files/publications/The%20Sedona%20Principles%20Third%20Edition.19TSCJ1.pdf.

The eDiscovery task is one of information retrieval, which has been a discipline of scientific study for more than 70 years.[4] Those 70 years have seen considerable advances in information retrieval methods, as well as ways to measure their effectiveness. One of the earliest and most primitive approaches to information retrieval is the so-called Boolean search, which retrieves only documents containing certain search terms or combinations of search terms, as specified by a manually constructed query. "Keyword culling" in eDiscovery almost always employs Boolean search, with minor enhancements that take into account the order and proximity of the terms within a document. Parties sometimes exchange "hit reports," quantifying the number of times the search terms appear in the document collection, but these reports provide little, if any, useful information about the quality of the search terms.

The most common—but certainly not the only—way to quantify the effectiveness of an information retrieval effort is to estimate recall and precision. These measures rely on the convenient fiction of binary relevance: that there is a "ground truth," and that any particular document either is or is not relevant to the information need that occasioned the information retrieval effort. If we could somehow know with certainty the relevance of every document in the collection to be searched, recall would be the percentage of all relevant documents that were retrieved, while precision would be the percentage of all retrieved documents that were relevant.[5] But we cannot know with certainty—or even with high confidence—the relevance of every document, or even the number of relevant documents in the collection, and thus recall. At best, we can estimate them with varying—and typically large—margins of error. Any method whose operation depends on a purportedly precise estimate of recall involves pseudo-science and bogus statistics.

While some leading information retrieval scientists eschew the use of recall altogether,[6] it is generally accepted that recall and precision can be used to gauge the *relative effectiveness* of competing information retrieval methods, provided that recall is measured under the same conditions, using the same information need, the same document collection, and the same independently derived relevance assessments.[7] The National Institute of Standards and Technology's ("NIST's") Text REtrieval Conference ("TREC") was founded in 1992, precisely to mount a heroic collaborative effort, among academic, industry, and government researchers, to create such conditions—conditions that cannot possibly be reproduced within the context of a particular eDiscovery effort.[8]  As such, efforts like TREC are useful to compare the effectiveness and reliability of different information retrieval methods, not to determine the effectiveness of any particular information retrieval effort.  Under TREC-like conditions, 65% recall and 65% precision represents a good result—approximately what we would expect if the entire collection were assessed by one subject matter expert ("SME"), with recall and precision estimated using the independent assessments of a second SME.[9]

---

[4] *See* Mark Sanderson and W. Bruce Croft. *The history of information retrieval research*, 100 Proc. of the IEEE 100 Special Centennial Issue 1444-51 (2012), https://ieeexplore.ieee.org/iel5/5/4357935/06182576.pdf.

[5] *See* Maura R. Grossman and Gordon V. Cormack, *The Grossman-Cormack Glossary of Technology-Assisted Review*, 7 Fed. Cts. L. Rev. 1, 25, 27 (2013), https://www.fclr.org/fclr/articles/html/2010/grossman.pdf.

[6] *See* Justin Zobel, Alistair Moffat, and Laurence A.F. Park, *Against recall: is it persistence, cardinality, density, coverage, or totality?*, in 43 ACM SIGIR Forum 3-8 (2009), https://dl.acm.org/doi/10.1145/1670598.1670600.

[7] *See* Ellen M. Voorhees, *Variations in relevance judgments and the measurement of retrieval effectiveness*, 36 Info. Processing & Mgmt. 697-716 (2000), ), https://doi.org/10.1016/S0306-4573(00)00010-8.

65% recall and 65% precision are not necessarily the best possible results.  If, for example, the majority vote of independent assessments by three SMEs were used, they might achieve on the order of 75% recall and 75% precision, according to the same evaluation conditions.[10]  For similar reasons, it is possible for certain specific TAR methods—where relevance is determined using input from both an SME *and* artificial intelligence—to achieve better effectiveness than a single SME alone.[11]

Much ink has been spilled over the entirely irrelevant question of whether or not Boolean search is a form of TAR.  Boolean search, like TAR, is an information retrieval method.  Boolean search using manually constructed queries (as, for example, employed for keyword culling in eDiscovery), unlike certain TAR methods, generally cannot achieve 70% recall and 70% precision.[12]  Perhaps one Boolean query can yield 70% recall and 20% precision, while another for the same information need can yield 20% recall and 70% precision.  But finding a single query that can yield both 70% recall and 70% precision is likely impossible.

To make matters worse, eDiscovery medicine shows frequently promote the sequential use of two or more information retrieval methods, including Boolean search, TAR, and manual review.  The net effect of this concoction is to achieve considerably lower recall than any of its constituent parts:  When multiple information retrieval methods are used in sequence, overall recall is the product of the recall for each constituent method.  If keyword culling were to achieve 70% recall, the TAR tool were to achieve 80% recall, and manual review were to achieve 75% recall, the recall of a review effort combining them in sequence would be 70%×80%×75%=42%. It is possible to quibble with the numbers presented here, but not with the fact that each constituent part is imperfect, and that overall or end-to-end recall is considerably less than the weakest link in the chain.[13]

---

[8] *See* Ellen M. Voorhees and Donna K. Harman (eds.), 63 *TREC:  Experiment and evaluation in information retrieval* (MIT Press, 2005).  *See also* TREC Proc., 1992–present, *available at* https://trec.nist.gov/proceedings/proceedings.html (last visited Sept. 28, 2021).

[9] *See* Ellen M. Voorhees, *supra* n.7.

[10] The majority vote of three independent assessments is more likely to agree with another independent assessment than any of the individual assessments.  Thus, when another independent assessment is used to evaluate recall and precision, both of these measures will likely be higher for the majority vote than for any individual.  This is an application of the well-known statistical phenomenon referred to as *regression to the mean*, "first noted by Sir Francis Galton that 'each peculiarity in man is shared by his kinsmen, but on the average to a less degree.'"  Brian S. Everett and Anders Skrondal, *The Cambridge Dictionary of Statistics, Fourth Edition* (Cambridge Univ. Press 2010), at 363, http://www.stewartschultz.com/statistics/books/Cambridge%20Dictionary%20Statistics%204th.pdf.

[11] *See* Gordon V. Cormack and Maura R. Grossman, *Navigating Imprecision in Relevance Assessments on the Road to Total Recall:  Roger and Me*, in Proc. of the 40th Int'l ACM SIGIR Conference on Research and Dev. in Info. Retrieval, 5-14 (2017), https://dl.acm.org/doi/10.1145/3077136.3080812; Maura R. Grossman and Gordon V. Cormack, *Technology-Assisted Review in E-Discovery Can Be More Effective and More Efficient Than Exhaustive Manual Review*, 17 Rich. J. L. & Tech. 1 (2011), https://scholarship.richmond.edu/cgi/viewcontent.cgi?article=1344&context=jolt.

[12] *See* Eero Sormunen, *Extensions to the STAIRS study—empirical evidence for the hypothesised ineffectiveness of Boolean queries in large full-text databases*, 4 Info. Retrieval 257-73 (2001), http://citeseerx.ist.psu.edu/viewdoc/summary?doi=10.1.1.24.1144.

When applied sequentially, information retrieval methods—whether Boolean search, TAR, or manual review—will always yield inferior recall. Yet the medicine shows would have us believe that we need to consider only the TAR-tool component in our recall calculations, ignoring relevant documents excluded by keyword culling and/or by post-TAR manual review. This is, at best, an extreme case of moving the goalposts, but more likely, a form of legerdemain.

No doubt purveyors and pundits will attack these arguments and examples with special pleading, *argumentum ad hominem*, appeals to common practice or Sedona Principle 6, strawmen, testimonials, and more pseudo-science—anything but a valid scientific experiment published in a peer-reviewed venue.

The following sections outline several of the methods of deception that have been used—and will continue to be used—to promote inferior methods and to cast aspersions on superior ones, until someone is ready to bell the cat.

**Misapplication of Effectiveness Measures**

Statistical measures are commonly used to evaluate the efficacy of medical treatments as well as information retrieval methods. A cancer treatment might be judged by its five-year survival rate, while an antibiotic might be judged by the probability that it will cure a specific infection. Neither of these statistics can possibly be measured on a case-by-case basis at the time of treatment; they are instead used to establish which treatments have the best chance of success, so that a reasonable choice can be made when the need arises, considering not only efficacy but cost, availability, side effects, and so on. On a case-by-case basis, observations like symptom abatement, temperature reduction, and the absence of a rash are indicators—but not proof—of successful treatment. These indicators of success often occur within a day or two of antibiotic treatment, but it generally takes several more days to eliminate the infection. If treatment is discontinued prematurely, the infection will likely return, in a form more resistant to the treatment than before. It would cause genuine harm to continue antibiotics only so long as symptoms persist, and then to repeat the treatment when symptoms recur. It might be possible to conduct a laboratory blood test to see with reasonable certainty when the infection was gone, but the test itself would entail more burden, cost, and delay than simply completing the prescribed course of antibiotics.

Recall and precision were conceived by scientists to measure the *comparative effectiveness* of information retrieval methods, not the success of a particular information retrieval effort, and especially not as the sole determinant of when a particular information retrieval effort may be discontinued. It is infeasible to compute a precise and accurate estimate of recall within the context of a particular

---

[13] *See* Maura R. Grossman and Gordon V. Cormack. *Comments on "The implications of Rule 26 (g) on the use of technology-assisted review,"* Fed. Cts. L. Rev. 285, 293-95 (2014), https://www.fclr.org/fclr/articles/pdf/comments-implications-rule26g-tar-62314.pdf.

eDiscovery effort,[14] and even if such an estimate were feasible, it would not inform the question of whether or not the review, if continued, would find enough additional relevant documents to justify the additional effort, a proportionality consideration that is included in the description of the scope of discovery set forth in Federal Rule of Civil Procedure 26(b)(1), and in the "reasonable inquiry" requirement of Federal Rule 26(g)(1).

The conflation of efficacy and success has led to the absurd propositions that no treatment or information retrieval method can be determined to be reasonable or unreasonable in advance, that no irreparable harm occurs when a treatment or search method fails in a particular case, and that success in a particular case can be solely determined by an estimate of a summary measure like body temperature or recall, without regard to symptoms or the quality of the production.

**Measuring the Wrong Quantity (a/k/a Searching Under the Streetlight)**

Recall and precision measure the *end-to-end effectiveness* of an information retrieval effort. As noted above, the scientific literature indicates that 70% to 80% recall is achievable by certain TAR methods. Many TAR methods employ a software tool in conjunction with manual review and/or prior Boolean keyword culling. The recall of the software tool, in isolation, must be considerably higher—perhaps 85% to 95%—to achieve an overall recall of 70% to 80% for the end-to-end search and review process, when correctly measured with respect to a blind independent assessment.

**Non-Blind Assessment**

Several of the most egregious applications of pseudo-science involve the use of non-blind experiments. Clever Hans was a horse purported to be able to perform intellectual tasks including arithmetic.[15] In fact, Hans' handler was either deliberately or inadvertently telegraphing the correct answer to Hans through non-verbal cues. A panel of 13 experts debunked the handler's claims by conducting a blinded experiment in which Hans was isolated from the questioner and spectators—Hans literally wore blinders—using questions for which the questioner did and did not know the answers in advance. The experiments concluded that Clever Hans could intuit the correct answer from the questioner, even if the questioner was not Hans' handler, but only if the questioner knew the answer in advance. So Clever Hans was actually less clever than initially thought.

Notwithstanding this debunking, the Clever Hans show continued to tour Germany, attracting large and enthusiastic crowds. Nowadays, recall is calculated from the non-blind relevance assessments of armies of Clever Hanses, to the delight of large and enthusiastic crowds at eDiscovery medicine shows.

**Misleading Statistics**

Statistics play several roles in information retrieval. First and foremost, statistics are used to calculate the efficacy and reliability of information retrieval methods over many information needs and/or document collections. Second, summary measures like recall and precision can be estimated for a particular information retrieval effort from relevance assessments for only a statistical sample of the collection. Third, many information retrieval methods rely on statistical algorithms to score and rank documents by their likelihood of relevance. While only the second use has any operational impact on eDiscovery, it is frequently conflated with the other two, and all are conflated with technology-assisted review.

---

[14] *See* David C. Blair, *STAIRS redux: Thoughts on the STAIRS evaluation, ten years after*, 47:1 J. Am. Soc. for Info. Sci. 4-22 (1996), https://asistdl.onlinelibrary.wiley.com/doi/abs/10.1002/%28SICI%291097-4571%28199601%2947%3A1%3C4%3A%3AAID-ASI2%3E3.0.CO%3B2-3.

[15] *See* Oskar Pfungst, *Clever Hans (The horse of Mr. Von Osten): A contribution to experimental animal and human psychology*," Translation by Carl. L Rahn (2010), *available at* https://www.gutenberg.org/files/33936/33936-h/33936-h.htm#CHAPTER_IV (last visited Sept. 28, 2021).

Measuring recall and precision—whether to establish the efficacy and reliability of a method or to assess the quality of a particular review—has little to do with the technology employed for the review, be it Boolean keyword culling and/or TAR and/or manual review. Regardless of what technologies are employed, recall is the fraction of relevant documents *in the entire collection* that are retrieved by the overall effort. On the other hand, the mechanics of statistical information retrieval algorithms are of no more concern to the eDiscovery practitioner than the biochemistry of a pharmaceutical is to the patient.

Part of the reason that these uses are conflated is that TAR was the first eDiscovery technique whose efficacy was evaluated and compared to that of manual review using the tools of information retrieval. TAR practitioners exposed to these methods naturally assumed that the use of statistics was unique to TAR, and that statistical expertise was needed to employ TAR. Purveyors and pundits have harnessed this misconception to suggest that only the TAR tool should be subject to validation, while keyword culling and manual review should be exempt, as they have always been.

At the same time, purveyors and pundits have grossly exaggerated the accuracy with which recall and precision (even of the TAR tool alone) can be estimated using sampling.[16] One tell that is often seen in vendor and pundit publications is the use of the non-sequitur "statistically significant sample." A more subtle mangling of statistical concepts is the notion of a sample having a given margin of error at a given confidence level; the most common examples are "margin of error of ±5% at a 95% confidence level," and "margin of error of ±2% at a 95% confidence level," when a more direct statement of what is really meant is "a random sample of size 385," and "a random sample of size 2,400," respectively. Specifying sample size obliquely by the margin of error that can purportedly be derived from it serves to obfuscate the sampling process and to intimidate practitioners, while strongly implying that the estimates derived from such samples are more precise than they are, and that they are sufficient to yield estimates of *recall* with the stated margins of error. They are not.

At best, samples of size 385—or even 2,400—can give a coarse estimate of recall with a margin of error several times larger than implied, which may identify a major failure, but cannot guarantee success. Furthermore, if the samples are assessed by Clever Hanses who know or can intuit how the documents were previously coded, we can place no bound on the error—and therefore the over- or underestimation—that can result.

**The Fallacies of Incredulity, Special Pleading, and Assuming the Converse**

Pundits' and practitioners' incredulity that large numbers of responsive documents are excluded by keyword culling and by reviewer assessment errors does not refute the extensive body of scientific evidence showing that they are. Nor does a special pleading argument that their own unique assessors or quality assurance process or bloodletting instruments would have achieved a different result. Nor does pointing out limitations—whether perceived or real—in scientific studies constitute evidence for the converse of the studies' conclusions. There is a strong body of scientific evidence showing that keyword culling removes a substantial number of responsive documents, that certain TAR methods achieve as good or better recall and precision than human assessors, and that non-blind assessments provide unreliable results. If any of the issues raised in arguments to the contrary had merit, they could surely be demonstrated through scientific studies whose results were subject to peer review. That has not happened.

---

[16] *See* Maura R. Grossman and Gordon V. Cormack *supra* n.13, at 305-10.

**The Ball is in The Courts' Court**

Just as bloodletting came to be considered unreasonable in the face of mounting scientific evidence, so too should certain common eDiscovery practices. Just as the snake oils of the early 20th century were unreasonable from the outset, so too are various TAR tools and methods whose efficacy has not been established.

What is needed is a full-scale evaluation, and judicial recognition of which methods are reasonable and which are not. Unfortunately, an evidentiary hearing sufficient to establish reasonableness would likely entail burden and cost disproportionate to the needs of any particular case, as evidenced by the *Kleen Products* case,[17] in which then-Magistrate (now retired) Judge Nan R. Nolan, after two full days of evidentiary hearings and 11 status hearings and Rule 16 conferences[18] with the parties, ordered them essentially to work it out for themselves.

In the near-decade since *Kleen Products*, the scientific body of evidence regarding how to determine the effectiveness and reliability of TAR methods has grown substantially. It is time to revisit what is reasonable in light of this evidence.

Federal Rule of Civil Procedure 26(g)(1) requires that the producing party or their attorney personally certify that "after reasonable inquiry," to the best of their knowledge, information, and belief, the production is not unreasonable, considering the burden or expense, the needs of the case, prior discovery in the case, the amount in controversy, and the importance of the issues at stake in the action.

Notwithstanding Sedona Principle 6's prescription that "[r]esponding parties are best situated to evaluate the procedures, methodologies, and technologies appropriate for preserving and producing their own electronically stored information," Federal Rule of Civil Procedure 26(g)(1)—which is the law—mandates that this evaluation must be consistent with a "reasonable inquiry," and that the chosen procedures, methodologies, and technologies must be calculated to yield a reasonable production.

The courts need not presume that any method used in the past or any method chosen by the producing party is reasonable, especially if presented with clear and convincing evidence that it is unreliable or ineffective, compared to readily available alternatives. In the face of such evidence, the court need not wait until the method's unreliability or ineffectiveness is manifested in the matter before it, which may occur only at the eleventh hour when proper remediation is no longer practical. At the same time, courts need not impose on producing parties the prescriptions of requesting parties unless they are similarly supported by clear and convincing evidence.

---

[17] *See Kleen Prods. LLC* v. *Packaging Corp. of Am.*, Civ. Ac. No. 10-C-5711, 2012 WL 4498465, at *4 (N.D. Ill. Sept. 28, 2012).
[18] *See* Fed. R. Civ. P. 16(a).

In assessing whether an inquiry, an eDiscovery method, or a production is reasonable, the courts may consider the *Daubert* factors:

> Many considerations will bear on the inquiry, including whether the theory or technique in question *can be (and has been) tested*, whether it *has been subjected to peer review and publication*, its *known or potential error rate*, and the *existence and maintenance of standards controlling its operation*, and whether it has *attracted widespread acceptance within a relevant scientific community*. (emphasis added).[19]

To be clear, we do not advocate a full-blown evidentiary hearing every time a TAR or validation process is challenged. Rather, we suggest that the *Daubert* factors offer useful guidance in determining what a reasonable process is pursuant to Fed. R. Civ. P 26(g)(1), and what is proper evidence to this effect.

The scientific literature contains a growing body of peer-reviewed and widely accepted evidence of the effectiveness and error rates of the information retrieval methods used in eDiscovery, including the error rates for Boolean search, manual review, specific TAR tools and protocols, and specific validation methods. eDiscovery and/or validation efforts can be considered tested only when the tools, methods, and standards of operation they employ are comparable to those of the research studies from which supporting evidence is derived.

Merely labeling an eDiscovery method as "TAR" does not establish that it has been tested. Nor does using a TAR tool that has been tested establish that it has been used according to a tested protocol. Nor does naming a particular protocol such as "Continuous Active Learning" or "CAL" establish that the protocol has been followed according to established standards of operation. Nor does proclaiming "75% recall" establish that recall has been calculated according to scientifically established methods with a known error rate, or that such a computation is evidence of either a reasonable inquiry or a reasonable production.

In contrast, producing parties should show—and the courts should demand that they show—the reasonableness of their eDiscovery search and review processes, as well as the resulting production, by hewing closely to tools, methods, and procedures that have been scientifically vetted and shown to be valid and reliable. Anything else belongs in a medicine show.

---

[19] *See* Syllabus (c) to *Daubert* v. *Merrell Dow Pharm., Inc.*, 509 U.S. 579, 580 (1993). The *Daubert* factors are incorporated in Federal Rule of Evidence 702 governing expert testimony. While it is unlikely that Federal Rule of Evidence 702 per se governs eDiscovery representations, there can be no question that the effectiveness and reasonableness of eDiscovery methods is a matter of scientific inquiry. Indeed, in *Victor Stanley, Inc.* v. *Creative Pipe, Inc.*, 250 F.R.D. 251, 260 n.10 (D. Md. 2008), then-Magistrate (now District) Judge Grimm noted that "[i]t cannot credibly be denied that resolving contested issues of whether a particular search and information retrieval method was appropriate—in the context of a motion to compel or motion for protective order—involves scientific, technical or specialized information. If so, then the trial judge must decide a method's appropriateness with the benefit of information from some reliable source—whether an affidavit from a qualified expert, a learned treatise, or, if appropriate, from information judicially noticed. To suggest otherwise is to condemn the trial court to making difficult decisions on inadequate information, which cannot be an outcome that anyone would advocate." We agree.